\documentclass[]{article}

\usepackage[title]{appendix}
\usepackage[affil-it]{authblk}
\usepackage[intlimits]{amsmath}
\usepackage{amsfonts}
\usepackage{wrapfig}

\usepackage{graphicx}
\usepackage{caption}

\usepackage{subfigure}
\usepackage[usenames]{color}
\usepackage{epstopdf,epsfig,todonotes}
\usepackage{xcolor}
\usepackage{indentfirst}
\usepackage{enumerate}
\usepackage{tikz}
\usepackage{mathtools}
\usepackage{graphicx}
\usepackage{amssymb, mathtools, mathrsfs}
\usepackage{comment}
\usepackage{upgreek}
\usepackage{hyperref}

\usepackage{tikz-3dplot}

\newcommand\be            {\begin{equation}}
	\newcommand\ee            {\end{equation}}

	\def\bea{\begin{equation}}
		\def\eea{\end{equation}}

	\newcommand\EN           {\end{equation}}
\newcommand\bes           {\begin{subequations}}
	\newcommand\esu           {\end{subequations}}

\newcommand\p            {\partial}

\renewcommand\p            {\partial}

\newcommand{\eref}{\eqref}

\newcommand{\bu}{\bi{u}}

\newcommand{\df}{\mathfrak{d}}

\newcommand{\bi}{\boldsymbol}

\usepackage{amsmath,amssymb}

\begin{document}

\title{Nonlinear Dynamics of Wind-Drift Currents at Mid-Latitudes}

	\author{Christian Puntini}
	\affil{Faculty of Mathematics, University of Vienna, Oskar-Morgenstern-Platz 1, 1090 Vienna,	Austria.}
	\maketitle

	\begin{abstract}
		Starting from the Navier-Stokes equation in the $f$-plane approximation, we provide an exact and explicit solution of the governing equations at leading order for fluid flows in the upper layer of the ocean at mid-latitudes, driven by a wind stress. Such a solution highlights the presence of a mean Ekman current superimposed to trochoidal oscillations and a background geostrophic current.
	\end{abstract}
	
	\noindent \textbf{Keywords:}  wind-driven currents, inertial oscillations, f-plane approximation, Ekman spiral, asymptotic methods.\\
	 \textbf{MSC:} 76M45, 76U60, 35B30, 35C05, 35Q30, 35Q35

	\vspace{0.5cm}


%
\tableofcontents

\section{Introduction}\noindent

The upper ocean, namely the topmost layer of sea water between the surface and the thermocline, with depths of the order of $50\,\mathrm{m}$ and rarely exceeding  $100\,\mathrm{m}$, is the region of the water column featuring the most complex dynamics. It is the primary stratum for the exchange of heat between the ocean and the atmosphere, as the first $2-3\, \mathrm{m}$ of the ocean posses the same heat capacity as the entire atmosphere above it. Moreover, around $50\%$ of the solar radiation is absorbed within the top $0.5\,\mathrm{m}$ of the ocean. Other energetic processes occur in this upper layer, such as wave breaking: about half of the breaking surface wave kinetic energy is dissipated within $20\%$ of the significant wave height from the surface (see \cite{SL}).  Also, this water stratum is highly affected by meteorological events, such as rain, contributing to the surface and volume of freshwater, momentum and heat fluxes. However, along with the physical processes cited above and many others (we refer to \cite{SL} and \cite{Talley} for a detailed description of the other phenomena occurring in the first meters of the ocean), there is another one that is particularly important: turbulence. In fact, the surface layer (also named surface mixing layer, as intensive mixing due to turbulence occurs there) is chaotic. Nevertheless, some coherent structures still emerge, such as wind-driven (Ekman) flows, and tidal and pressure-driven (geostrophic) currents.\\
Ekman flows (or Ekman spirals), named after V. W. Ekman, who in 1905 was the first to provide a quantitative explanation for the phenomenon, describe the deflection of steady wind-driven ocean currents to the right of the prevailing wind direction in the Northern Hemisphere. This deflection arises from the momentum balance between the Coriolis acceleration, which acts on the upper ocean layers, and the frictional forces generated by turbulent stress induced by the wind \cite{Ekman1905}. The deflection angle is to the left in the Southern Hemisphere.\\
Ekman's classical theory predicts a deflection angle of  $45^{\circ}$; however, \textit{in situ} measurements have not only confirmed the existence of the Ekman spiral but also highlighted significant deviations between the observed deflection angles and those predicted by the classical solution (see \cite{Polton}, \cite{Roach}, \cite{YM}, \cite{poulain} and \cite{RC}).\\
In this paper, we extend the classical Ekman solution to investigate wind-driven nonlinear flows in the upper ocean layer, focusing on mid-latitude regions—areas away from the Equator and excluding the poles, studying the nonlinear governing equations, at leading order, for wind-driven surface currents. A similar analysis for the Arctic Ocean has been conducted in \cite{C2022a}, \cite{C2022b}, and \cite{mio}. We remark that these solutions are not applicable to equatorial regions, as the Coriolis force vanishes at the Equator and its sign reversal across the Equator generates azimuthal (i.e., longitudinal) flows (see, e.g., \cite{CI} and \cite{CJ2015}).\\
Our approach involves the application of systematic theoretical methods to address problems in physical oceanography, following the framework established by works such as \cite{CJ2016}, \cite{J2022}, and \cite{JBook}. While it is undeniable that data are essential to support both new ideas and existing theories, we contend that valuable insights into the dynamics of geophysical flows can still be achieved through the careful (asymptotic) analysis of the equations of motion and the boundary conditions that define the problem (see also \cite{J2017} and \cite{J2019} for a review of this methods).\\
The general approach for this theoretical framework is the following: after having identified the typical scales of the fluid flows under consideration, the full set of differential equations, boundary and initial conditions are written in non-dimensional variables and some non-dimensional parameters are identified. Usually, at least one of these parameters is small. Then, one proceed to construct an asymptotic solution using the small parameter as the basis for the asymptotic expansion. \\
A final remark is in order. The flows considered in this study are characterized by length scales that are small compared to the Earth’s dimensions. Specifically, as will be discussed in detail later, they are relevant to regions with spatial extents of approximately \(1^{\circ}\) in latitude or longitude. Consequently, it is reasonable as a first approximation to employ a local Cartesian coordinate system and to formulate the Navier–Stokes equations under the \(f\)-plane approximation \cite{Vallis}. It should be noted, however, that the \(f\)-plane approximation has inherent limitations, as it is derived by neglecting certain \emph{ad hoc} terms in the full equations of motion expressed in spherical coordinates, and by adopting a tangent-plane Cartesian framework (see \cite{CJ2019a} and \cite{Vallis}). The extension of the present analysis to a more general formulation will be addressed in future work.\\
\\
\noindent
This paper is organized as follows. In Section \ref{sec:governing}, after introducing the equations of motion for a viscous fluid together with appropriate characteristic scales, the governing equations are non-dimensionalised, and a single small parameter, \(\varepsilon\), defined as the ratio between the vertical length scale and the Earth's radius, is introduced and used as basis for the asymptotic expansion. All other parameters are held fixed, i.e. \(\mathcal{O}(1)\), as we take the limit \(\varepsilon \to 0\) to derive the leading-order governing equations. The boundary conditions are similarly non-dimensionalised and analysed in the limit \(\varepsilon \to 0\).\\
In Section \ref{sec-Lag} we present an exact solution to the nonlinear governing equations---at leading order---adopting the Lagrangian description of fluid motion, by specifying the trajectories of the individual fluid particles. These paths are composed of a mean Ekman spiral, trochoidal time-dependent near-inertial oscillations and a background geostrophic current, thus extending the classical solution of Ekman \cite{Ekman1905} to include also near-inertial flow, that is relevant in the large-scale ocean dynamics (see \cite{ripa}, \cite{XV} and \cite{WF}). Moreover, we prove that, given the wind velocity and the background geostrophic current, the wind-drift surface current will be uniquely determined by the boundary conditions, with a deflection angle from the blowing wind—measured relative to the geostrophic flow—of less than $45^{\circ}$; this result is compatible with some measurements (see \cite{poulain} and \cite{YM}). \\
For simplicity, we will analyse the fluid motion in the Northern Hemisphere, but the results apply immediately in the Southern Hemisphere, where the deflection angle is to the left.\\
Finally, the results are discussed in Section \ref{sec-discussion}.

\section{Governing equations}\label{sec:governing}
\begin{figure}[ht!]
\centering
    \includegraphics[width=0.5\linewidth]{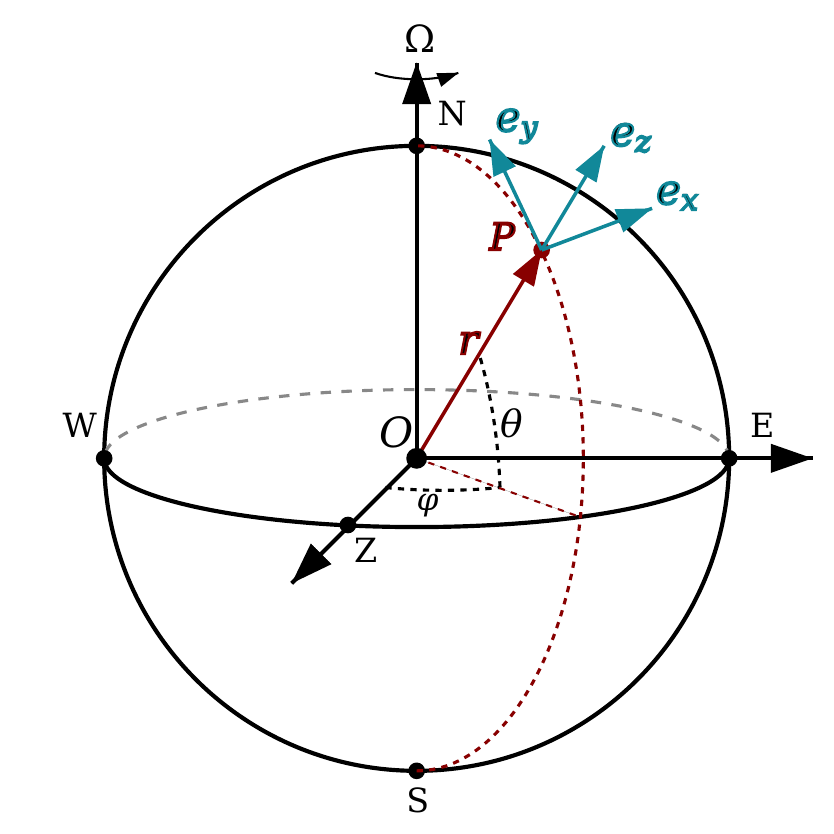}
    \caption{The Cartesian coordinate system on the rotating Earth. Here, $\bf{r}$ represents the position vector of a point located at latitude $\theta$ and longitude $\varphi$. Z represents the null island.}
    \label{fig-spherical}
\end{figure}

\noindent
We introduce a local Cartesian coordinate system with unit vectors $(\mathbf{e}_1, \mathbf{e}_2, \mathbf{e}_3)$ and coordinates $(x', y', z')$. The $(x', y', z')$-system is associated with a point fixed on the sphere (other than at the two poles where $\mathbf{e}_1$ and $ \mathbf{e}_2$ are not well-defined) which is rotating about its polar axis (with constant angular speed $\Omega' \approx 7.29\cdot10^{-5}\, \mathrm{rad\, s^{-1}}$). See Figure \ref{fig-spherical}. The associated velocity components are $(u',v',w')=:\bu'$. \\
We use prime to denote physical (dimensional) quantities; the prime will be removed after introducing the appropriate set of non-dimensional variables.\\
We adopt the $f$-plane approximation, as we are considering flows with small length scales (compared to Earth's dimensions), hence the Coriolis parameters $f'=2\Omega'\sin\theta $ and $\hat{f}'=2\Omega'\cos\theta$ are considered constant, with $\theta$ denoting the fixed angle of latitude relative to the sector of the ocean being analyses. The local tangential coordinates $x'$ and $y'$, pointing eastward and northward, respectively, will describe the position of points within the small regions under consideration, for every fixed $z'$. The coordinate $z'$ describes the distance from the Earth's centre in the radial direction.  As the flows we are studying here are those present in the topmost stratum of the ocean, we anticipate that $z'$ will be rewritten as $z'=R'+\zeta'$, where $R' \approx 6371\, \mathrm{km}$ is the radius of the Earth (considered a sphere of constant geopotential).\\
The Navier-Stokes equation---in the $f$-plane approximation--- is therefore written as (see \cite{Vallis} or \cite{W02}),
\begin{equation}\label{NS 0}
    \frac{D\bu'}{D t'}+(\hat{f}' w'-f'v',\,f'u', -\hat{f}'u')=\\
 =-\frac{1}{\rho'(z)}\nabla' p'+ (0,0,-g')+A_{\mathrm{h}}' \left(\frac{\p^2}{\p x'^2}+\frac{\p^2}{\p y'^2}\right)\bu'+A_{\mathrm{v}}' \frac{\p^2}{\p z'^2}\bu' 
\end{equation}
where $A'_{\mathrm{h}}$ and $A'_{\mathrm{v}}$ are the horizontal and vertical kinematic eddy viscosities, respectively; they are assumed to be constant, and we have taken the density to be only depth-dependent, that is $\rho'=\rho'(z)$, which is common for oceanic flows (see e.g. \cite{CJ2023}). In \eref{NS 0} we have used the common shorthand notation for the gradient and for the material derivative 
\begin{equation}
      \nabla'= \left( \frac{\partial}{\partial x'},\  \frac{\partial}{\partial y'} ,\   \frac{\partial}{\partial z'}
      \right),\quad
        \frac{D}{Dt'}=\frac{\partial} {\partial t'}+ u'\frac{\partial}{\partial x'}+v' \frac{\partial}{\partial y'} +w'  \frac{\partial}{\partial z'},
\end{equation}
where $t'$ is the time variable.
The equation of mass conservation is 
\begin{equation}\label{mass}
    \frac{D\rho'}{D t'}+\rho'\left(\frac{\p u'}{\p x'}+\frac{\p v'}{\p y'}+\frac{\p w'}{\p z'}\right)=0,
\end{equation}
which, due to the assumption on the viscosity (implying $\frac{D\rho'}{D t'}=w'\frac{d\rho'}{dz'}$) reduces to

\begin{equation}\label{mass 1}
    \frac{w'}{\rho'}\frac{d\rho'}{dz'}+\frac{\p u'}{\p x'}+\frac{\p v'}{\p y'}+\frac{\p w'}{\p z'}=0.
\end{equation}
As anticipated, we write  $z'=R'+\zeta'$, with $R'$ being the Earth's radius, and we simplify the Navier-Stokes equation \eref{NS 0} by redefining the pressure as
\begin{equation}
    p'=P_{\mathrm{atm}}+g'\int_{z'}^{R'}\rho'(r)\, dr+ P'(x',y',z', t'),
\end{equation}
leading to
\begin{equation}\label{NS 1}
    \frac{D\bu'}{D t'}+(\hat{f}' w'-f'v',\,f'u', -\hat{f}'u')=-\frac{1}{\rho'(z)}\nabla' P'+A_{\mathrm{h}}' \left(\frac{\p^2}{\p x'^2}+\frac{\p^2}{\p y'^2}\right)\bu'+A_{\mathrm{v}}' \frac{\p^2}{\p z'^2}\bu'.
\end{equation}
 Now we non-dimensionalise the equations of motion \eref{NS 1} and \eref{mass 1} by introducing new variables $x,y,z,u,v,w,\rho,P$  according to the transformation
\begin{equation}\label{scaling}
\begin{aligned}
&(x',y')=L'(x,y),\quad \zeta'=D'z, \quad  t'=\frac{L'}{U'}t,\\
 &  (u',v',w')=U'(u,v,k w),\quad \rho'=\overline{\rho'} \rho, \quad P'=\overline{\rho'} U'^2P.
\end{aligned}
   \end{equation}
Here $L'$ is a length scale, $D'$ is a suitable depth scale (to be chosen later), $U'$ an appropriate speed scale, $\overline{\rho'}$ is the average density and $k$ a scaling factor.\\
We assume the length scale $L'$ to be of the order---but not in excess---of the Rossby radius of deformation $\mathcal{L}'(\theta)$. We report here the Rossby radius at four latitudes (see \cite{Chelton})
\begin{equation}\label{RR}
    \mathcal{L}'(10^{\circ})\approx 100\,\mathrm{km},\quad \mathcal{L}'(20^{\circ})\approx 50\,\mathrm{km}, \quad \mathcal{L}'(45^{\circ})\approx 30\,\mathrm{km},\quad \mathcal{L}'(60^{\circ})\approx 20\,\mathrm{km}.
\end{equation}
We introduce the ``thin-shell" parameter
\begin{equation}
    \varepsilon=\frac{D'}{R'},
\end{equation}
and we denote by $\upmu$ the ratio between $L'$ and $R'$,
\begin{equation}
    \upmu=\frac{L'}{R'}.
\end{equation} 
Because the length scales we are considering are small (cf. \eref{RR}), the regions we are considering will not exceed $1^{\circ}$ of latitude. As a consequence, the $f$-plane approximation emerges as a suitable initial choice for the study of such problems, even if it possesses some limitations (see the discussion in \cite{CJ2019a}). We hope, in a subsequent work, to extend this kind of analysis to a more general setting.\\
After the non-dimensionalisation \eref{scaling}, the Navier-Stokes equation \eref{NS 1} reads as (in component form)
\begin{equation}\label{primo scaling}
		\left\{	\begin{aligned}
	\frac{\partial u}{\partial t} + u \frac{\partial u}{\partial x} + v \frac{\partial u}{\partial y} + k\frac{L'}{D'}w\frac{\partial u }{\partial z} 
-fv + fk w \cot\theta= -\frac{1}{\rho(z)} \frac{\partial P}{\partial x}
+ \frac{L'^2}{D'^2} \frac{1}{\mathrm{Re_{\mathrm{v}}}} \frac{\partial^2 u}{\partial z^2}
+ \frac{1}{\mathrm{Re_{\mathrm{h}}}} \left( \frac{\partial^2 u}{\partial x^2} + \frac{\partial^2 u}{\partial y^2} \right), \\  
  \frac{\partial v}{\partial t} + u \frac{\partial v}{\partial x} + v \frac{\partial v}{\partial y} + k\frac{L'}{D'}w\frac{\partial v}{\partial z} 
+ f u = -\frac{1}{\rho(z)} \frac{\partial P}{\partial y}
+ \frac{L'^2}{D'^2} \frac{1}{\mathrm{Re_{\mathrm{v}}}} \frac{\partial^2 v}{\partial z^2}
+ \frac{1}{\mathrm{Re_{\mathrm{h}}}} \left( \frac{\partial^2 v}{\partial x^2} + \frac{\partial^2 v}{\partial y^2}  \right), \\  
  k\left(\frac{\partial  w}{\partial t} + u \frac{\partial  w}{\partial x} + v \frac{\partial w}{\partial y} + k\frac{L'}{D'}w\frac{\partial  w}{\partial z} \right)
- f u \cot\theta=-\frac{1}{\rho(z)} \frac{L'}{D'} \frac{\partial P}{\partial z}
+ \frac{L'^2}{D'^2} \frac{k}{\mathrm{Re_{\mathrm{v}}}} \frac{\partial^2 w}{\partial z^2}
+ \frac{k}{\mathrm{Re_{\mathrm{h}}}} \left( \frac{\partial^2 w}{\partial x^2} + \frac{\partial^2 w}{\partial y^2}   \right),
\end{aligned}\right.
\end{equation}
where we have introduced the non-dimensional parameters
\begin{equation}
    f=2\frac{\Omega' L'}{U'}\sin\theta, \qquad\mathrm{Re_{\mathrm{h}}}=\frac{U'L'}{A_{\mathrm{h}}'},  \qquad\mathrm{Re_{\mathrm{v}}}=\frac{U'L'}{A_{\mathrm{v}}'}.
\end{equation}
The first of these is an inverse Rossby number and the other two are Reynolds numbers. \\
With the same scaling \eqref{scaling}, the equation of mass conservation \eqref{mass 1} becomes 
\begin{equation}\label{mass 2}
  \frac{\p u}{\p x}+\frac{\p v}{\p y}+k\frac{L'}{D'}  \left(\frac{w}{\rho}\frac{d\rho}{dz}+\frac{\p w}{\p z}\right)=0.
\end{equation}
Following \cite{CJ2019a}, we make use of $\frac{L'^2}{D'^2 \mathrm{Re_{\mathrm{v}}}}$ to define the relevant depth-scale $D'$ by setting
\begin{equation}\label{D'def}
D'=\sqrt{\frac{L' A'_{\mathrm{v}}}{ U'}} ,
\end{equation}
giving
\begin{equation}
  \frac{L'^2}{D'^2}= \mathrm{Re_{\mathrm{v}}}\qquad\text{and}\qquad  \frac{1}{\mathrm{Re_{\mathrm{h}}}}=\frac{D'^2}{L'^2}\upnu=\frac{\varepsilon^2}{\upmu^2}\upnu,
\end{equation}
where
\begin{equation}
    \upnu:=\frac{A'_{\mathrm{h}}}{A'_{\mathrm{v}}}=\frac{\mathrm{Re_{\mathrm{v}}}}{\mathrm{Re_{\mathrm{h}}}}.
\end{equation}
This choice for the depth-scale $D'$ is related to the fact that wind-drift currents are due to the balance between the Coriolis force and the viscous stress (hence they have to be of the same order). For reference, we set the value of the mean vertical kinetic eddy viscosity (also referred to as the coefficient of vertical turbulent exchange, see \cite{K77}) to be
\begin{equation}\label{eddyV}
    A_{\mathrm{v}}'\approx0.01\, \mathrm{m^2s^{-1}}.
\end{equation}
A value for the horizontal kinematic viscosity is difficult to obtain, but it is possible to say that ${A'_{\rm h}} \gg {A'_{\rm v}}$, with their ratio $\upnu$ typically of order $10^3$ (see \cite{Pedlosky, Talley}).\\
Assuming $U'\approx0.1\, \mathrm{ms^{-1}}$, which is the typical magnitude of ocean currents (see e.g. \cite{CJ2019b}), and considering \eref{RR} and \eref{eddyV}, we find that the values of $D'$ are in the range 
\begin{equation}\label{D'}
  45\,\mathrm{m}  \lesssim D'\lesssim 100\,\mathrm{m},
\end{equation}
which are, in fact, the standard depths of the near-surface layer of the ocean, where wind-drift effects are significant (see \cite{K77}), validating our choice of the depth scale $D'$ in \eref{D'def}.\\
Moreover, it follows that $f\approx 25$, if we use $L'=\mathfrak{L'}(\theta)$ at the specific latitudes \eref{RR}.\\
The equations of motion \eref{primo scaling} and \eref{mass 2} therefore becomes
\begin{equation}\label{secondo scaling}
	\left\{
	\begin{aligned}
  \frac{\partial u}{\partial t} + u \frac{\partial u}{\partial x} + v \frac{\partial u}{\partial y} + \kappa \upmu w \frac{\partial u}{\partial z}
-fv + f\kappa \varepsilon w \cot\theta 
=-\frac{1}{\rho(z)} \frac{\partial P}{\partial x}
+ \frac{\partial^2 u}{\partial z^2}
+ \frac{\varepsilon^2}{\upmu^2} \upnu \left( \frac{\partial^2 u}{\partial x^2} + \frac{\partial^2 u}{\partial y^2}  \right),  \\
  \frac{\partial v}{\partial t} + u \frac{\partial v}{\partial x} + v \frac{\partial v}{\partial y} + \kappa \upmu w \frac{\partial v}{\partial z} 
+ f u
=-\frac{1}{\rho(z)} \frac{\partial P}{\partial y}
+ \frac{\partial^2 v}{\partial z^2}
+ \frac{\varepsilon^2}{\upmu^2} \upnu \left( \frac{\partial^2 v}{\partial x^2} + \frac{\partial^2 v}{\partial y^2}  \right), \\
  \frac{\kappa \varepsilon^2}{\upmu} \left( \frac{\partial w}{\partial t} + u \frac{\partial w}{\partial x} + v \frac{\partial w }{\partial y} + \kappa \upmu w \frac{\partial w}{\partial z} \right) 
- \frac{\varepsilon}{\upmu}f u \cot\theta =-\frac{1}{\rho(z)} \frac{\partial P}{\partial z}
+ \frac{\kappa \varepsilon^2}{\upmu} \frac{\partial^2 w}{\partial z^2}
+ \frac{\kappa \varepsilon^4}{\upmu^3} \upnu \left( \frac{\partial^2 w}{\partial x^2} + \frac{\partial^2 w}{\partial y^2}  \right),
\end{aligned}\right.
\end{equation}
and
\begin{equation}\label{mass 3}
  \frac{\p u}{\p x}+\frac{\p v}{\p y}+\kappa\upmu  \left(\frac{w}{\rho}\frac{d\rho}{dz}+\frac{\p w}{\p z}\right)=0.
\end{equation}
\\
After having non-dimensionalised the governing equations, we aim to provide a solution describing the fluid flow under consideration, but, as it is well known, there are no explicit solutions to the Navier-Stokes equations. Therefore, we seek an approximate solution capturing the most important features of the physical problem we are analysing.\\
To this end, we write an asymptotic expansion based on the ``thin-shell" parameter $\varepsilon$, that is, denoting any of the unknowns $u,v,w,P$ and $\rho$ by $\mathsf{q}$, we write
\begin{equation}\label{asymptotic} 
\mathsf{q}(x,y,z,t;\varepsilon) = \sum_{j\geq 0}\varepsilon^j \mathsf{q}_{j}(x,y,z,t) .\end{equation}
The choice of $\varepsilon$ as base for the asymptotic expansion reflects the fact that the flows under consideration have small depth-scales compared to the radius of the Earth.\\
Moreover, another characteristic of wind-driven currents is that the coupling between the vertical and the horizontal motion is very weak (see e.g., \cite{CJ2019a}); therefore we  elect to consider the class of problems for which, in the limit $\varepsilon\rightarrow 0$,
\begin{equation}
    k=\varepsilon\kappa \qquad \text{with} \qquad \kappa=o(1).
\end{equation}
As an example, we could set $\kappa = \varepsilon^n$, with $n > 0$, as proposed in \cite{CJ2024} in the context of atmospheric flows, and then we may construct a complete asymptotic solution based on $\varepsilon$.  However, in this work we will only present the leading-order solution (hence, the exact definition of $\kappa$ is not crucial, but it suffices to assume that $\kappa = o(1)$ in the limit $\varepsilon \to 0$).
At leading order---that is, in the limit $\varepsilon \to 0$, with $\kappa = o(1)$ and keeping the other parameters fixed---the momentum equations \eref{secondo scaling} take the form:
\begin{equation}\label{governing enqs}
\left\{\begin{aligned}	\frac{\partial u}{\partial t} + u \frac{\partial u}{\partial x} + v\frac{\partial u}{\partial y} - fv= -\frac{1}{\rho}\frac{\partial P}{\partial x} + \frac{\partial^2 u}{\partial z^2},\\
	\frac{\partial v}{\partial t} + u \frac{\partial v}{\partial x} + v\frac{\partial v}{\partial y} +fu= -\frac{1}{\rho}\frac{\partial P}{\partial y} +\frac{\partial^2 v}{\partial z^2},\\
\frac{\p P}{\p z}=0.   \end{aligned}\right. 
	\end{equation}
  We remark the fact that the limit $\varepsilon\rightarrow 0$ is not a real limit in the mathematical sense, but rather a formal way to obtain the equations at leading order ($j=0$).\\
    The continuity equation \eref{mass 3} at leading order simplifies as
\begin{equation}\label{1cont}
	\frac{\partial u}{\partial x} + 	\frac{\partial v}{\partial y}=0.
\end{equation}
From \eref{governing enqs} and \eref{1cont}, the flow appear to be a depth-dependent $2D$ horizontal flow.\\
Note that, since we are focusing exclusively on the order $j = 0$, the subscript $0$ has been omitted in equations \eqref{governing enqs} and \eqref{1cont}. For simplicity, we write $u$ instead of $u_0$, and similarly for $v$, $w$, and $P$. This notation will also be used in the derivation of the boundary conditions at leading order, as well as in the computation of the leading-order solution throughout the remainder of this paper.

\subsection{Boundary conditions}
\begin{figure}[ht!]
    \centering
    \includegraphics[width=\linewidth]{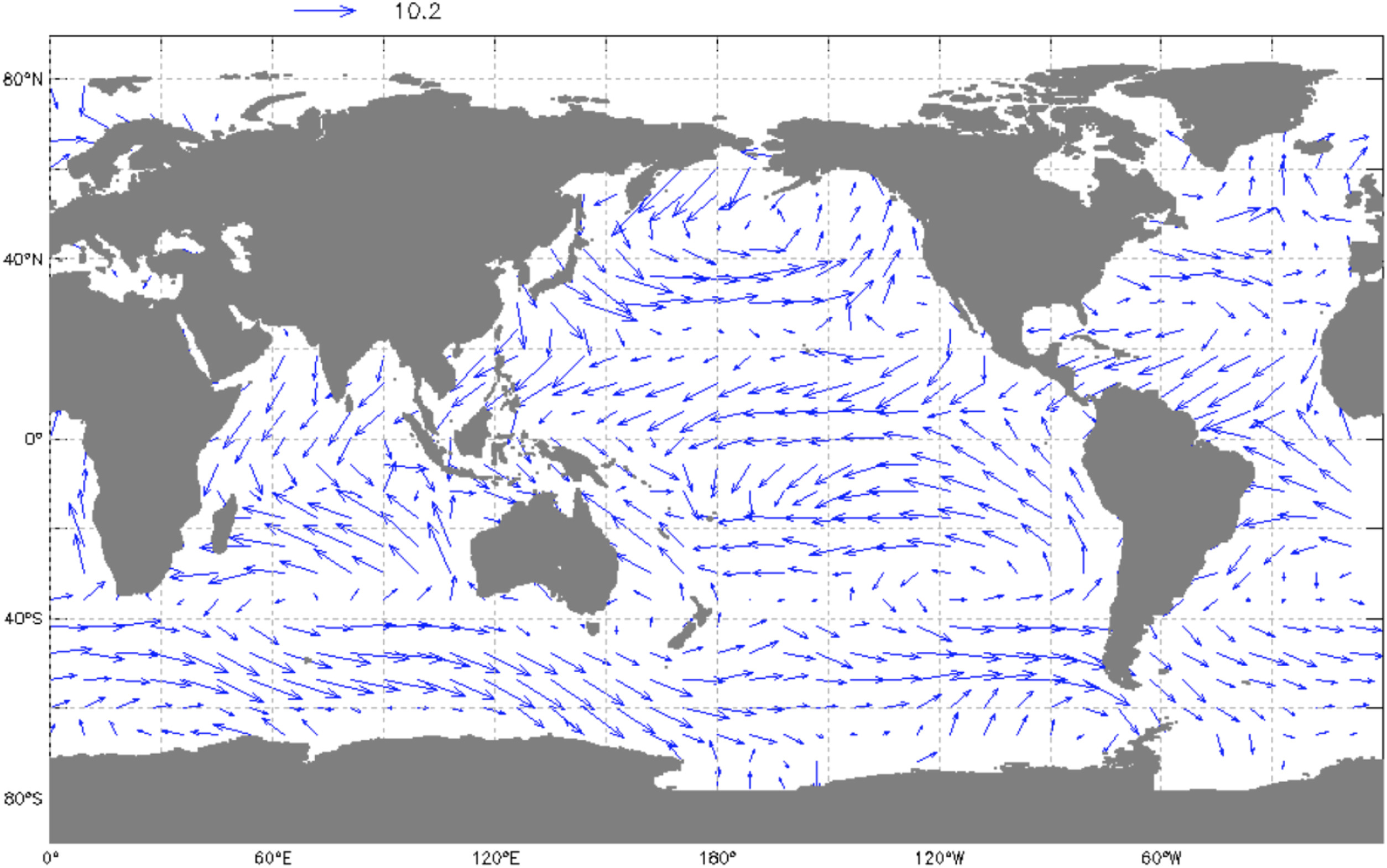}
    \caption{Monthly-averaged ocean wind speed and direction vectors, with vector lengths proportional to the reference scale (in $\mathrm{m\, s^{-1}}$), based on observations from NASA's QuikSCAT satellite. Image credit: NOAA.}
    \label{wind}
\end{figure}

\noindent
For the full characterization of the geophysical flow under consideration, a major role is played by the boundary conditions. We impose the following boundary conditions at the free surface $\zeta'=h'(x',y',t')$ (recall that we are writing $z'=R'+\zeta'=R'+D'z$):
\begin{itemize}
    \item dynamic boundary condition: we impose a constant surface pressure, namely
    \begin{equation}\label{dynamicBX}
        p'=P_{\mathrm{atm}}\qquad  \text{on} \qquad \zeta'=h';
    \end{equation}
    \item kinematic boundary condition: we assume that no fluid particles will leave the fluid via the free surface, namely
    \begin{equation}\label{KBC}
        w'=\frac{\p h'}{\p t'}+u'\frac{\p h'}{\p x'}+v'\frac{\p h'}{\p y'}\qquad  \text{on} \qquad \zeta'=h';
    \end{equation}
    \item shear stress: we require the shear stress at the surface of the water to be given. Denoting by $\rho'_{\rm s}$ the water density at the surface, we express this in the form
    \begin{equation}\label{stress1}
           \tau_1'(x',y', t')=\rho'_{\rm s}A'_{\mathrm{v}}\frac{\p u'}{\p z'},\quad
           \tau_2'(x',y', t')=\rho'_{\rm s}A'_{\mathrm{v}}\frac{\p v'}{\p z'},
      \qquad  \text{on} \qquad \zeta'=h';
       \end{equation}
       \item depth-dependence: we require that the wind-drift currents decrease as depth increases. As the vertical velocity component is neglected at leading order in \eref{governing enqs} and \eref{1cont}, the boundary condition capturing the decay with depth of wind drift flows is given by
\begin{equation}\label{21}
\frac{	\partial (u'^2+v'^2)}{\partial z'}<0\qquad  \text{for} \qquad z'<h'.
\end{equation}
\end{itemize}
The boundary conditions also need to be non-dimensionalised according to \eref{scaling}, and, analogously to the equations of motion, studied in the limit $\varepsilon\rightarrow 0 $ to retain the leading-order terms.\\
We begin from the kinematic boundary condition \eref{KBC}, which, under the scaling \eqref{scaling}, reads as
\begin{equation}\label{KBC adim}
        \kappa\varepsilon w=\frac{H'}{L'}\left(\frac{\p h}{\p t}+u\frac{\p h}{\p x}+v\frac{\p h}{\p y}\right)\qquad  \text{on} \qquad \zeta'=h',
    \end{equation}
    where we have written $h'=H'h$. As the depth scale $H'$ is independent of $\varepsilon$, in the limit $\varepsilon\rightarrow0$, \eref{KBC adim} reduces to
    \begin{equation}\label{KBC adim 2}
        \frac{\p h}{\p t}+u\frac{\p h}{\p x}+v\frac{\p h}{\p y}=0\qquad  \text{on} \qquad z=\frac{H'}{D'}h.
    \end{equation}
    The condition \eref{KBC adim 2} implies the free surface is invariant for the flow, so the most reasonable physical assumption is that the free surface is flat. We can assume, without loss of generality, that the flat free surface is given by $z=0.$\\
    Note that the presence of a flat free surface---namely the absence of surface waves---emerges as a consequence of the asymptotic behaviour of the governing equations due to the scaling \eref{scaling} relevant to the problem we are considering.  In particular, surface waves arise if $u'$ and/or $v'$, and $w'$ are of the same order (see \cite{SL}), while a fundamental assumption of the flow we are considering is that the vertical velocity $w'$ has order of magnitude much lower that the horizontal velocities $u'$ and $v'$, by an order of $\kappa\varepsilon$.\\
    Due to the approximation of the free surface as a flat one, the condition \eref{dynamicBX} became
\begin{equation}\label{7new}
	P=0\qquad  \text{on} \qquad z=0,
\end{equation}
while the boundary condition \eqref{21} capturing the depth-dependence of the currents become
\begin{equation}\label{decrease}
\frac{	\partial (u^2+v^2)}{\partial z}<0\qquad  \text{for} \qquad z<0,
\end{equation}
when non-dimensionalised. At leading order, the boundary conditions \eqref{7new} and \eqref{decrease} remain the same.\\
The stress generated by the wind blowing over the water is given by the bulk formula
\begin{equation}\label{stress2}
    \bi{\tau}'_{\rm a}=\rho'_{\rm a} C_{\mathrm{aw}} |\bu'_{\rm a}-\bu'|(\bu'_{\rm a}-\bu') \qquad  \text{on} \qquad \zeta'=h',
\end{equation}
where $\rho'_{\rm a}$ is the air density, $C_{\mathrm{aw}}$ is the (non-dimensional) drag coefficient of air over water and $\bu'_{\rm a}= (u'_{\rm a},v'_{\rm a})$ is the wind velocity describing the horizontal momentum exchange at the surface.\\
The boundary condition \eqref{stress2} express the fact that, if there is a current present at the sea surface, the only contribution to the stress generated by the wind is given by the velocity of the wind relative to the ocean surface current, and not just by the wind velocity in its entirety. \\ In the literature, the wind velocity is usually referred as the wind velocity at $10\,\mathrm{m}$ above the sea surface, and, for order of magnitude estimates we follow \cite{SL} and set
\begin{equation}\label{caw}
C_{\mathrm{aw}} \approx 
\begin{cases}
1.1 \cdot 10^{-3}, & \text{for } |\mathbf{u}'_{\mathrm{a}}| \lesssim 5\,\mathrm{m\,s^{-1}}, \\[6pt]
(0.61 + 0.063|\mathbf{u}'_{\mathrm{a}}|) \cdot 10^{-3}, & \text{for } 5\,\mathrm{m\,s^{-1}} \lesssim |\mathbf{u}'_{\mathrm{a}}| \lesssim 22\,\mathrm{m\,s^{-1}},
\end{cases}
\end{equation}
the second one being the linear relationship proposed by \cite{S88}. For winds blowing over $22\,\mathrm{m\,s^{-1}}$, a formula for the drag coefficient is difficult to obtain (see \cite{SL}). However, this is not a big problem, as these are extreme wind conditions that eventually occur rarely and for short periods (usually these are gusts); as we are considering a time scale of the order of some days (cf. \eqref{scaling} and \eqref{RR}), wind gusts over short time-scales are not relevant for our analysis. See Figure \ref{wind} for order of magnitude of the blowing wind at different latitudes.  The relation \eqref{stress2} coupled with \eqref{caw} is consistent with the observations of a mean wind stress of magnitude $0.01$--$0.1~\mathrm{N\,m^{-2}}$ over the ocean surface (see the data in \cite{Paldor2024} and \cite{Talley}).\\
Equating \eref{stress1} and \eref{stress2}, namely asking the continuity of the shear stress across the water surface, gives
\begin{equation}\label{stress3}
   \rho'_{\rm s}A'_{\mathrm{v}}\left(\frac{\p u'}{\p z'},\, \frac{\p v'}{\p z'}\right) =\rho'_{\rm a} C_{\mathrm{aw}} \sqrt{(u_{\rm a}'-u')^2+(v_{\rm a}'-v')^2}\,(u'_{\rm a}-u',\,v'_{\rm a}-v')\qquad  \text{on} \qquad \zeta'=h'.
\end{equation}
Writing $(u'_{\rm a},\,v'_{\rm a})=U'(u_{\rm a},v_{\rm a})$, the non-dimensional form of \eref{stress3} reads as
\begin{equation}\label{stressADIM prima}
    \left(\frac{\p u}{\p z},\, \frac{\p v}{\p z}\right) =\gamma_{\mathrm{aw}} C_{\mathrm{aw}} \sqrt{(u_{\rm a}-u)^2+(v_{\rm a}-v)^2}\,(u_{\rm a}-u,\,v_{\rm a}-v)\qquad  \text{on} \qquad z=0,
\end{equation}
where we have defined the non-dimensional parameter
\begin{equation}
    \gamma_{\mathrm{aw}}=\frac{\rho'_{\rm a}\, U'\, D'}{\rho'_{\rm s} A'_{\mathrm{v}}}.
\end{equation}
Using the Taylor expansion $\sqrt{1+x} = 1 + \frac{1}{2}x + o(x)$ as $x\to 0$, and writing, according to \eqref{asymptotic} $u=u_0+\varepsilon u_1+o(\varepsilon)$, $v=v_0+\varepsilon v_1+o(\varepsilon)$ we see that
\begin{equation} \label{stress expanded}
\begin{aligned}
\sqrt{(u_{\rm a}-u)^2+(v_{\rm a}-v)^2} = \sqrt{(u_{\rm a}-u_0)^2+(v_{\rm a}-v_0)^2} 
 - \varepsilon\frac{u_1(u_{\rm a} - u_0) + v_1(v_{\rm a} - v_0)}{\sqrt{(u_{\rm a}-u_0)^2+(v_{\rm a}-v_0)^2}} + o(\varepsilon) \quad \text{on} \quad z=0.
\end{aligned}
\end{equation}
At leading order, the boundary condition \eqref{stressADIM prima} therefore reads as (dropping the subscripts $0$ as before)
\begin{equation}\label{stressADIM}
    \left(\frac{\p u}{\p z},\, \frac{\p v}{\p z}\right) =\gamma_{\mathrm{aw}} C_{\mathrm{aw}} \sqrt{(u_{\rm a}-u)^2+(v_{\rm a}-v)^2}\,(u_{\rm a}-u,\,v_{\rm a}-v)\qquad  \text{on} \qquad z=0.
\end{equation}
\section{The nonlinear solution in Lagrangian framework}\label{sec-Lag}
\noindent
In this section we provide a leading-order solution to the nonlinear wind-drift problem given by the differential equations \eqref{governing enqs} and \eqref{1cont} and the boundary conditions \eqref{KBC adim 2}, \eqref{7new}, \eqref{decrease} and \eqref{stressADIM}.\\
We start by decomposing the horizontal velocity field into its geostrophic components $\left(U_{\mathrm{g}}, V_{\mathrm{g}}\right)$ and ageostrophic components $\left(\widehat{u}, \widehat{v}\right)$. This decomposition is consistent with the ocean being in geostrophic balance, where the presence of a pressure gradient is balanced by a geostrophic current (see \cite{Kundu} and \cite{R2025}). We therefore write
\begin{equation}\label{uv}
    u=U_{\mathrm{g}}+\widehat{u},\qquad v=V_{\mathrm{g}}+\widehat{v},
\end{equation}
with the geostrophic velocity components satisfying
\begin{equation}\label{geoP}
		\frac{\partial P}{\partial x}=\rho f\,V_{\mathrm{g}},\qquad
	\frac{\partial P}{\partial y}=-\rho f\,U_{\mathrm{g}}, 
	\end{equation}
    and we assume $U_{\mathrm{g}}$ and $V_{\mathrm{g}}$ to be constant. Due to \eref{uv} and \eref{geoP},
the first two equations of \eref{governing enqs},  reduces to
\begin{equation}\label{53}
		\frac{D u}{D t}  - f\widehat{v}= \frac{\partial^2 u}{\partial z^2},\qquad
		\frac{D v}{D t}+f\widehat{u}=  \frac{\partial^2 v}{\partial z^2},
	\end{equation}
while the third one, namely 	$\frac{\partial P}{\partial z}=0$ remains unchanged.\\
Adopting the Lagrangian approach (that is describing the motion of every single particle of fluid), and specifying at every time $t$ the positions
\begin{equation}\label{Lag}
\left\{	\begin{aligned}
		x(t; a,b,z)=a + \left[d_1(z)+U_{\mathrm{g}}\right]t -\frac{1}{k} e ^{k(b+z)}\sin(k(a-z-ct)), \\
			y(t; a,b,z)=b + \left[d_2(z)+V_{\mathrm{g}}\right]t +\frac{1}{k} e ^{k(b+z)}\cos(k(a-z-ct)) ,
        \end{aligned}\right.
	\end{equation}
of the horizontally moving fluid particles in terms of the depth $z$, the material variables $a,b$ and the parameters $k>0$ and $c>0$, for suitably chosen functions $d_1(z)$ and $d_2(z)$, we will show that a solution to the nonlinear wind-drift flow \eref{governing enqs}, \eref{1cont} satisfying the boundary conditions is given by \eref{Lag}. For the labeling variables, we have that $a\in\mathbb{R}$, while $b\in(b_1,b_0)$, with $b_1<b_0<0$, ensuring an exponential decay for $z\leq 0$ in \eref{Lag}.\\ Writing
\begin{equation}\xi=k(a-z-ct),\end{equation}
it follows that at every fixed time $t$ we have
\begin{equation}\label{jac1}
\begin{pmatrix}
\frac{\partial x}{\partial a} & \frac{\partial x}{\partial b} \\
\frac{\partial y}{\partial a} & \frac{\partial y}{\partial b}
\end{pmatrix}=\begin{pmatrix}
1 - e^{k(b+z)}\cos(\xi) & -e^{k(b+z)}\sin(\xi) \\
-e^{k(b+z)}\sin(\xi) & 1 + e^{k(b+z)}\cos(\xi)
\end{pmatrix},
\end{equation}
with inverse given by
\begin{equation}
\begin{pmatrix}
\frac{\partial a}{\partial x} & \frac{\partial a}{\partial y} \\
\frac{\partial b}{\partial x} & \frac{\partial b}{\partial y}
\end{pmatrix}
=
\frac{1}{1 - e^{2k(b+z)}}
\begin{pmatrix}
1 + e^{k(b+z)}\cos(\xi) & e^{k(b+z)}\sin(\xi) \\
e^{k(b+z)}\sin(\xi) & 1 - e^{k(b+z)}\cos(\xi)\end{pmatrix}.
\end{equation}
The determinant of the matrix in \eref{jac1} is equal to $1-e^{2k(b+z)}$, hence the flow is area-preserving, and the continuity equation \eref{1cont} is satisfied (see \cite{CBook}). \\
From \eref{Lag} we have that
\begin{equation}\label{Lag u v}
\left\{\begin{aligned}
	u&=\frac{\partial x}{\partial t}= d_1(z)+U_{\mathrm{g}} +c e ^{k(b+z)}\cos(\xi),\\
	v&=\frac{\partial y}{\partial t}= d_2(z)+V_{\mathrm{g}} +ce ^{k(b+z)}\sin(\xi),
    \end{aligned}\right.
	\end{equation}
giving, for the ageostrophic components of the velocity field,
\begin{equation}\label{ageoV}
\left\{\begin{aligned}
	\widehat{u}&= d_1(z)+c e ^{k(b+z)}\cos(\xi),\\
	\widehat{v}&= d_2(z)+ce ^{k(b+z)}\sin(\xi).
    \end{aligned}\right.
	\end{equation}
Moreover, since
\begin{equation}\label{Jac2}
\begin{aligned}
\begin{pmatrix}
\frac{\partial u}{\partial x} & \frac{\partial v}{\partial x} \\
\frac{\partial u}{\partial y} & \frac{\partial v}{\partial y}
\end{pmatrix}
&=
\begin{pmatrix}
\frac{\partial a}{\partial x} & \frac{\partial b}{\partial x} \\
\frac{\partial a}{\partial y} & \frac{\partial b}{\partial y}
\end{pmatrix}
\begin{pmatrix}
    \frac{\partial u}{\partial a} & \frac{\partial v}{\partial a} \\
\frac{\partial u}{\partial b} & \frac{\partial v}{\partial b}
\end{pmatrix} \\
&=
\frac{kc}{1 - e^{2k(b+z)}}
\begin{pmatrix}
-e^{k(b+z)}\sin(\xi) & e^{k(b+z)}\cos(\xi) + e^{2k(b+z)} \\
e^{k(b+z)}\cos(\xi) - e^{2k(b+z)} & e^{k(b+z)}\sin(\xi)
\end{pmatrix},
\end{aligned}
\end{equation}
at each horizontal level the two-dimensional horizontal flow \eref{Lag} has constant vorticity
\begin{equation}\label{vorticity}
	\frac{\partial u}{\partial y}- \frac{\partial v}{\partial x}=\frac{2kc }{1-e^{-2k(b+z)}}.
	\end{equation}
    Using the expressions
\begin{equation}\label{Lag du dv}
\left\{\begin{aligned}
&\frac{D u}{D t}=\frac{\partial u}{\partial t} + u \frac{\partial u}{\partial x} + v\frac{\partial u}{\partial y} =kc^2 e ^{k(b+z)}\sin(\xi),\\
	&	\frac{D v}{D t}=\frac{\partial v}{\partial t} + u \frac{\partial v}{\partial x} + v\frac{\partial v}{\partial y} =-kc^2e ^{k(b+z)}\cos(\xi),
        \end{aligned}\right.
	\end{equation}
from \eref{53} we get the system of differential equations
\begin{equation}\label{32}
\left\{\begin{aligned}
f d_2(z)+  d_1''(z)
    +\left(fc + 2   k^2 c - k c^2\right)e^{k(b+z)}\sin(\xi)=0,\\
	f d_1(z)-  d_2''(z)      +\left(fc + 2   k^2 c - k c^2\right)e^{k(b+z)}\cos(\xi)=0.
    \end{aligned}\right.
	\end{equation}
Given the dispersion relation 
\begin{equation}\label{34}
c=\frac{f}{k}+2  k,
\end{equation} 
and writing in complex-variable notation
\begin{equation}
d(z)=d_1(z) + i d_2(z),
\end{equation}
 the system \eref{32} reduces to
\begin{equation}\label{ODE}
d''(z)-if d(z)=0,
\end{equation}
whose general solution is given by
\begin{equation}
	d(z)=A\, e^{(1+i)\lambda z}+ B\,e^{-(1+i)\lambda z},
\end{equation}
where $\lambda=\sqrt{\frac{f}{2 }}$ and $A, B \in \mathbb{C}$.   \\
As wind drift currents are insignificant at great depth, namely imposing the boundary condition \eref{decrease}, we have that $B=0$. The solution of \eref{ODE} therefore takes the form
\begin{equation}\label{38}
d(z)=d(0)e^{(1+i)\lambda z},
\end{equation}
with $d(0)=d_1(0)+id_2(0)\in \mathbb{C}$. The time-average of the velocity \eref{Lag u v} over a period $T=\frac{2\pi}{ck}=\frac{2\pi}{f+2 k^2}$, which in view of \eref{38} reads as
\begin{equation}  \label{velocity}
u+iv=d(0)e^{(1+i)\lambda z} +c e^{k(b+z)}e^{ik(a-z-ct)}+\bi{U}_{\mathrm{g}},
\end{equation}
where $\bi{U}_{\mathrm{g}}=U_{\mathrm{g}}+iV_{\mathrm{g}}$, yields the mean drift current
\begin{equation} \label{meancurrent2}
\langle u+iv\rangle_T=d(0)e^{(1+i)\lambda z}+\bi{U}_{\mathrm{g}}.
\end{equation} 
Here $\langle\cdot\rangle_T$ represents the time-average, and from \eref{meancurrent2} it is evident that $d(0)$ is the mean wind-drift ocean current at the surface.\\
Due to \eref{meancurrent2}, the boundary condition \eref{stressADIM} reduces---in an averaged sense---to 
\begin{equation}\label{402}
\frac{\lambda  }{C_{\mathrm{aw}}\gamma_{\mathrm{aw}}} \left(\df_1-\df_2, \df_1+\df_2\right)= |(u_{\rm a}-u,v_{\rm a}-v)|(u_{\rm a}-u,v_{\rm a}-v),
\end{equation}
in which the unknowns are $\df_1=d_1(0)$ and $\df_2=d_2(0)$. Given that $d(0)$ is the mean wind-drift ocean current at the surface, we set $u=U_{\mathrm{g}}+d_1(0),\ v=V_{\mathrm{g}}+d_2(0)$, and \eref{402} can be rewritten as (in complex notation and defining  $\beta=\frac{C_{\mathrm{aw}}\gamma_{\mathrm{aw}}} {\sqrt{2} \lambda } $)
\begin{equation}\label{drift2}
 (\df_1+i\df_2)=e^{-i\frac{\pi}{4}}\beta\left|(u_{\rm a}-U_{\mathrm{g}}-\df_1)+i(v_{\rm a}-V_{\mathrm{g}}-\df_2)\right|\left[(u_{\rm a}-U_{\mathrm{g}}-\df_1)+i(v_{\rm a}-V_{\mathrm{g}}-\df_2)\right].
\end{equation}
 Writing 
\begin{equation}\label{71}
    (u_{\rm a}-U_{\mathrm{g}}-\df_1)+i(v_{\rm a}-V_{\mathrm{g}}-\df_2)=\mathcal{U}e^{i\Phi},
\end{equation}
the boundary condition \eref{drift2} reads as
\begin{equation}\label{BD 1}
 \mathcal{U}^2e^{i\Phi}+ \frac{1+i}{\beta}\mathcal{U}e^{i\Phi}=\frac{Z}{\beta},
\end{equation}
where 
\begin{equation}\label{Z}
    Z=(u_{\rm a}-U_{\mathrm{g}}-v_{\rm a}+V_{\mathrm{g}}) + i (u_{\rm a}-U_{\mathrm{g}}+v_{\rm a}-V_{\mathrm{g}}).
\end{equation}
For $Z\not=0$, from \eref{BD 1} we get that
\begin{equation}\label{62}
 e^{i\Phi}=\frac{Z}{\mathcal{U}(\beta\mathcal{U}+ (1+i))},
\end{equation}
and, multiplying \eref{62} by its complex conjugates gives the polynomial expression
\begin{equation}\label{43}
   \mathcal{P}(\mathcal{U})= \beta^2\mathcal{U}^4+2\beta\mathcal{U}^3+2\mathcal{U}^2-|Z|^2=0.
\end{equation}
As $\mathcal{P}(0)<0$, $\lim_{R\rightarrow\pm \infty} \mathcal{P}(R)=+\infty$ and $\mathcal{P}''(R)>0$, $\mathcal{P}$ is strictly convex, ensuring that there is one solution $\mathcal{U}>0$ of \eref{43}. Therefore, we point out that such computations have highlighted how the boundary condition \eref{stressADIM} determines in a unique way the wind-drift surface current $d(0)$, knowing the geostrophic current and the wind speed. More precisely, due to \eref{71} and \eref{Z}, it is necessary just to know $u_{\rm a}-U_{\mathrm{g}}$ and $v_{\rm a}-V_{\mathrm{g}}$ to determine the wind-drift current at the surface, $d(0)$.\\
Writing the speed of the wind relative to the geostrophic current as
\begin{equation}
    (u_{\rm a}-U_{\mathrm{g}})+i(v_{\rm a}-V_{\mathrm{g}})=\mathcal{W}e^{i\Theta}
\end{equation}
and 
\begin{equation}
    d(0)=\df_1+i\df_2=\mathcal{D}e^{i\Psi}
\end{equation}
we can rewrite the expression \eref{drift2} as
\begin{equation}\label{drift3}
(1+i)\mathcal{D}e^{i\Psi}=\sqrt{2}\beta\left|\mathcal{W}e^{i\Theta}-\mathcal{D}e^{i\Psi}\right|\left(\mathcal{W}e^{i\Theta}-\mathcal{D}e^{i\Psi}\right)
\end{equation}
or, equivalently,
\begin{equation}\label{drift4}
\frac{(1+i)\mathcal{D}}{\sqrt{2}\beta\left|\mathcal{W}e^{i\Theta}-\mathcal{D}e^{i\Psi}\right|}+\mathcal{D}=\mathcal{W}e^{i(\Theta-\Psi)}.
\end{equation}
The angle $\Theta-\Psi$ represents the deflection angle between the wind velocity (relative to the oceanic geostrophic current) and the wind-drift flow at the surface. As the left side of \eref{drift4} is a complex number with imaginary part positive and smaller than the real part, it follows that $0<\Theta-\Psi<\pi/4$. See Figure \ref{Angolo} \\
\begin{figure}[ht]
    \centering
\begin{tikzpicture}[>=stealth, scale=2]

    \draw[->] (-0.5,0) -- (3,0) node[right] {$u$};
    \draw[->] (0,-0.5) -- (0,3) node[above] {$v$};

    \draw[->,very thick,red] (0,0) -- (1,2.3) node[midway, above left] {wind};
    \draw[->,very thick,blue] (0,0) -- (0.5,0.2) node[midway, below right] {$\bi{U}_{\mathrm{g}}$};
    \draw[->,very thick, olive] (0.5,0.2) -- (1,2.3) node[midway, above]{} ;
    \draw[->,very thick,black] (0.5,0.2) -- (1.8,1.52) node[right] {$d(0)$};

    \draw[thick, dashed] (1.2,0.9) arc[start angle=15, end angle=68, radius=0.7];
    \node at (1.2,1.5)  {$\Theta-\Psi$};
\end{tikzpicture}
\caption{Schematic depiction of the surface currents: in red the wind velocity; in blue the geostrophic current, in black the surface current with the angle $\Theta-\Psi$ to the right of the vector of the wind velocity relative to the geostrophic current (olive green) }
\label{Angolo}

\end{figure}
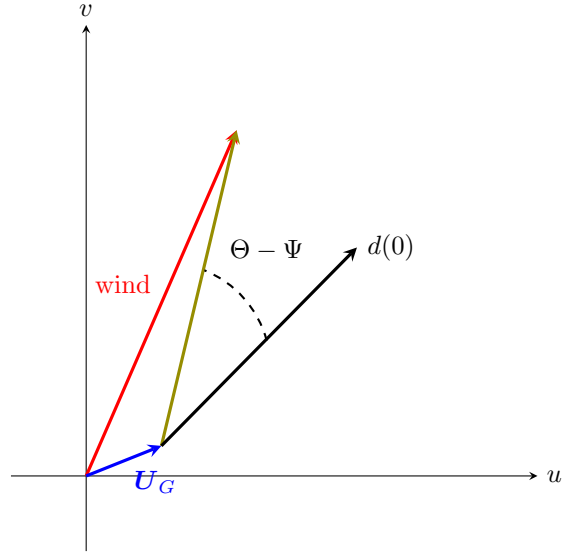




  

\begin{figure}
    \centering
  
\tdplotsetmaincoords{70}{50} 

\begin{tikzpicture}[tdplot_main_coords, scale=3]

    \draw[->] (-0.5,0,0) -- (2,0,0) node[right] {$u$}; 
    \draw[->] (0,-0.5,0) -- (0,2,0) node[above] {$v$}; 
    \draw[->] (0,0,0) -- (0,0,-2.2) node[below left] {$z$}; 

    \draw[->,very thick,blue] (0,0,0) -- (0.5,0.2,0) node[above] {$\bi{U}_{\mathrm{g}}$};
    \draw[->,very thick,black] (0.5,0.2,0) -- (1.8,1.52,0) node[right] {$d(0)$};

    \draw[domain=-2:0, smooth, variable=\t, samples=200, very thick, red] 
        plot ({0.5+ exp(5*\t) * (1.3*cos(deg(5*\t)) - 1.32 * sin(deg(5*\t)))}, 
              {0.2 + exp(5*\t) * (1.3*sin(deg(5*\t)) + 1.32 * cos(deg(5*\t)))},
              {\t}); 

     \foreach \t in {-0.8,-0.6,-0.5,-0.4,-0.3,-0.2,-0.1,-0.05} {
        \draw[->, thick, violet] 
            (0.5,0.2,\t) -- 
            ({0.5+ exp(5*\t) * (1.3*cos(deg(5*\t)) - 1.32 * sin(deg(5*\t)))}, 
             {0.2 + exp(5*\t) * (1.3*sin(deg(5*\t)) + 1.32 * cos(deg(5*\t)))},
             {\t});
    }

     \draw[dashed,thick,purple]  (0.5,0.2,0)-- (0.5,0.2,-2);
\end{tikzpicture}
\caption{The wind-drift current. In blue the geostrophic current, in black the wind-driven surface current, and in red the Ekman spiral. The $u, v$-axes, and the $z$-axis are not to scale. }
    \label{Ekman 3d}
\end{figure}
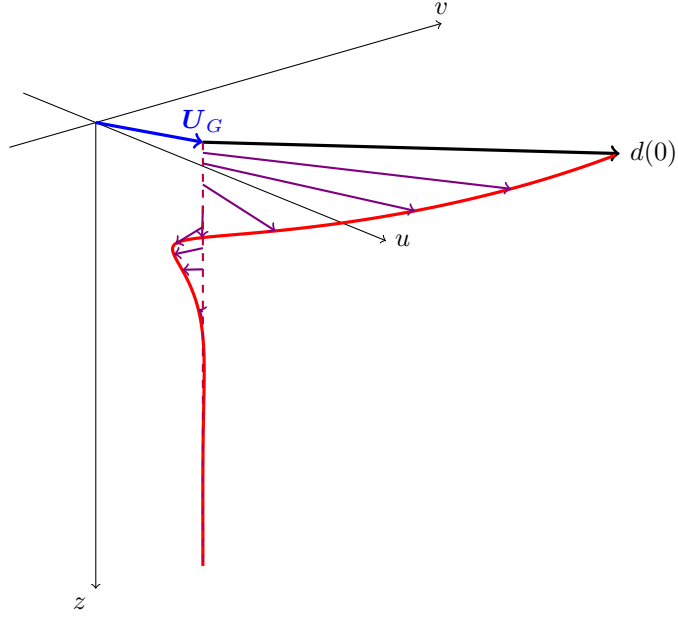

Of particular interest is also the structure of the solution \eref{Lag}. In fact, for any given $k>0$, the horizontal particle path given by \eref{Lag} coupled with \eref{38}
\begin{equation}\label{44}
    x(t;a,b,z)+iy(t;a,b,z)=(a+ib)+\left[d(0)e^{(1+i)\lambda z}+\bi{U}_{\mathrm{g}}\right]t+\\
   +\frac{1}{k}e^{k(b+z)}e^{i\left(\frac{\pi}{2}+k(a-z)-(f+2 k^2)t\right)},
\end{equation}
is a solution of \eref{governing enqs} and \eref{1cont} satisfying the boundary conditions and representing parametric trochoids. The mean ageostrophic current, over a period $T=\frac{2\pi}{ck}=\frac{2\pi}{f+2 k^2}$,
\begin{equation}\label{45}
\langle \widehat{u}+i\widehat{v}\rangle_T=\frac{1}{T}\int_0^T\left( \widehat{u}(t;a,b,z)+i\widehat{v}(t;a,b,z)\right) dt=d(0)e^{(1+i)\lambda z}
	\end{equation}
is the classical Ekman spiral (see Figure \ref{Ekman 3d}), while the oscillatory perturbation of the current 
\begin{equation}
\frac{1}{k}e^{k(b+z)}e^{i\left(\frac{\pi}{2}+k(a-z)-(f+2 k^2)t\right)}
\end{equation}
is near-inertial, as  $ f\approx 25$ and $k<<1$ (see \cite{C2022a}, \cite{mio}) gives
\begin{equation}
f+2 k^2\approx f
\end{equation}
for the frequency, hence the period of oscillations is approximately the inertial period of the Earth. \\
Averaging over the depth the expression in \eref{45} gives the depth-averaged mean-drift Ekman current of the solution \eref{44}, also known as (time-averaged) Ekman trasport,
\begin{equation}\label{48}
\mathcal{I}_{Ek}:=\int_{-\infty}^0 d(0))e^{(1+i)\lambda z}\, dz=\frac{d(0)}{\lambda \sqrt{2}}e^{-i\frac{\pi}{4}},
\end{equation}
which is at $45^{\circ}$ to the right of the mean surface current $d(0)$; this is a classical result of Ekman's theory.

\section{Discussion}\label{sec-discussion}\noindent
The presented analysis provides an explicit solution for the nonlinear governing equations---at leading order---for the wind-driven flows in regions away from the poles (where classical spherical coordinates fails, and the use of rotated spherical coordinates is necessary---see \cite{CJ2023} and \cite{mio}) and the equatorial band (due to the breakdown of the Ekman flow---see \cite{CJ2015}). The solution is the sum of a mean wind-drift current (the well-known Ekman spiral), a background geostrophic current balancing the pressure gradient, and near-inertial oscillations. For every fixed value of $z$, the particle path described \eref{44} represents a trochoid, the only known solution for the $2D$ incompressible Euler equations with free boundary discovered by Gerstner (see e.g. \cite{Milne}, \cite{Henry}, \cite{CS} or \cite{AP}). This study extends the solutions of \cite{C2022a}, \cite{C2022b} and \cite{mio} for the Arctic Ocean, where ice and the Transpolar Drift Current are present, to mid-latitudes regions. \\
Recalling the momentum equations at leading order
\begin{equation}
\left\{\begin{aligned}
		\frac{D u}{D t}  - fv= -\frac{1}{\rho}\frac{\partial P}{\partial x} +  \frac{\partial^2 u}{\partial z^2},\\
		\frac{D v}{D t}+fu= -\frac{1}{\rho}\frac{\partial P}{\partial y} +  \frac{\partial^2 v}{\partial z^2},
        \end{aligned}\right.
	\end{equation}
we can think the solution \eref{44} as a nonlinear ``superposition'' of two terms, namely the Ekman spiral, solution of
\begin{equation}
\left\{\begin{aligned}
		 - fv=   \frac{\partial^2 u}{\partial z^2},\\
		fu=  \frac{\partial^2 v}{\partial z^2},
                \end{aligned}\right.
	\end{equation}
and the Gerstner's trochoidal solution of the Euler equations 
\begin{equation}
\left\{\begin{aligned}
		\frac{D u}{D t}  = -\frac{1}{\rho}\frac{\partial P}{\partial x},\\
		\frac{D v}{D t}= -\frac{1}{\rho}\frac{\partial P}{\partial y} .
        \end{aligned}\right.
	\end{equation}
    We also have shown that, knowing the velocity of the wind and of the background geostrophic current, the wind-drift surface current is uniquely determined by the boundary conditions. Moreover, the deflection angle between the wind-drift flow at the surface and the blowing wind (relatively to the geostrophic flow) is less than $45^{\circ}$. This result on the deflection angle matches the measurements conducted in the Eastern Mediterranean \cite{poulain} and in the Tsushima strait \cite{YM}.\\
In this sense, we also point out a the importance of the boundary conditions. In some textbooks or papers (see e.g. \cite{TO} or \cite{MP}) the stress of the wind over the ocean is expresses as 
\begin{equation}\label{stress4}
    \bi{\tau}'_{\rm a}=\rho'_{\rm a} C_{\mathrm{aw}} |\bu'_{\rm a}|\bu'_{\rm a},
\end{equation}
without taking into account the surface motion of the ocean. Assuming \eref{stress4}, equation \eref{stressADIM} would be
\begin{equation}
    \left(\frac{\p u}{\p z},\, \frac{\p v}{\p z}\right) =\gamma_{\mathrm{aw}} C_{\mathrm{aw}} |(u_{\rm a},v_{\rm a})|\,(u_{\rm a},\,v_{\rm a}),
\end{equation}
which, due to  \eref{meancurrent2}, reads as
\begin{equation}\label{4021}
\frac{\lambda  }{C_{\mathrm{aw}}\gamma_{\mathrm{aw}}} \left(\df_1-\df_2, \df_1+\df_2\right)= |(u_{\rm a},v_{\rm a})|\,(u_{\rm a},v_{\rm a}),
\end{equation}
or equivalently as
\begin{equation}\label{drift 5}
(\df_1+i\df_2)=e^{-i\frac{\pi}{4}}\beta\sqrt{u^2_{\rm a}+v^2_{\rm a}}\,(u_{\rm a}+iv_{\rm a}),
\end{equation}
where we have written again $\df_1=d_1(0)$, $\df_2=d_2(0)$ and $\beta=\frac{C_{\mathrm{aw}}\gamma_{\mathrm{aw}}} {\sqrt{2} \lambda } $. In this case, the deflection angle would be of exactly $45^{\circ}.$\\
We have assumed the vertical eddy viscosity to be constant, which is, to a first approximation, a good choice for oceanic flows; however, a direction that is of interest is to extend this model to the case of depth-dependent or time-dependent eddy viscosities, expanding the recent investigations by \cite{BC}, \cite{CDP}, \cite{DPC}, \cite{C2021}, \cite{R2021} and \cite{R2022}). Observations indicate that eddy viscosity exhibits temporal variability on both daily \cite{WMcP16} and monthly timescales \cite{RF}, as well as depth-dependent profiles (see e.g. \cite{1} and \cite{2}), which in turn affects the characteristics of the Ekman spiral. \\
A limitation of our analysis lies in the assumption of a flat surface, as the effects of surface waves were not considered. This approximation allowed us to simplify the problem and compute an explicit solution. However the drift in the uppermost ocean layer is not only driven by currents but is also influenced by surface waves (see e.g. \cite{rascle2009} and \cite{rohrs2012}); also, in the first few meters the effects of the so-called wave-induced Stokes drift are present (see \cite{stokes1847}, \cite{monismith2008} and \cite{rohrs2014wave}).
Moreover, the dissipation of wave momentum (see \cite{carniel2009} and \cite{christensen2009}) and its interaction  with Earth’s rotation---through the Coriolis–Stokes force---can alter the structure and direction of surface currents (we refer to \cite{ursell1950theoretical}, \cite{Hasselmann}, \cite{lewis2004} and \cite{PoltonETAL}).\\
Accounting for all these processes in model considered will, surely, improve the accuracy of the result, but paying the price of strongly increasing the complexity of the equations, leading to, probably, a system that can not be solved explicitly. \\
We conclude by highlighting one essential feature of the explicit solution resulted from our analysis: the presence of near-inertial oscillations. These are not captured by the classical Ekman theory, but is known that they play a significant role in the slow, large-scale dynamics of the ocean (see \cite{XV}, \cite{ripa} and \cite{inertia}). 
\section*{Acknowledgments}
\begin{itemize}
\item  The author is grateful to the referees for their useful comments and suggestions.
    \item The author declares no conflict of interest.
    \item The author is supported by the Austrian Science Fund (FWF) [grant number Z 387-N, grant doi: https://doi.org/10.55776/Z387].
\end{itemize}

\end{document}